\documentclass[11pt]{article}
\usepackage{amsfonts}
\usepackage{amsmath}
\usepackage{amssymb}
\usepackage{indentfirst}
\usepackage{graphicx}

\begin{document}

\title{On the Supersymmetric Extension of Gauss-Bonnet like Gravity}
\author{P. K. Concha$^{1,2}$\thanks{%
patillusion@gmail.com}, M. C. Ipinza$^{3,4,5}$\thanks{%
marcelo.calderon@polito.it}, L. Ravera$^{4,5}$\thanks{%
lucrezia.ravera@polito.it}, E. K. Rodr\'{\i}guez$^{1,2}$\thanks{%
everodriguezd@gmail.com}, \\
{\small $^{1}$\textit{Departamento de Ciencias, Facultad de Artes Liberales, }}\\
{\small \textit{Universidad Adolfo Ib%
\'{a}\~{n}ez,}}\\
{\small Av. Padre Hurtado 750, Vi\~{n}a del Mar, Chile}\\
{\small $^{2}$\textit{Instituto de Ciencias F\'{\i}sicas y Matem\'{a}ticas,
Universidad Austral de Chile,}}\\
{\small Casilla 567, Valdivia, Chile}\\
{\small $^{3}$\textit{Departamento de F\'{\i}sica, Universidad de Concepci%
\'{o}n,}} \\
{\small Casilla 160-C, Concepci\'{o}n, Chile}\\
{\small $^{4}$\textit{DISAT, Politecnico di Torino}}\\
{\small Corso Duca degli Abruzzi 24, I-10129 Torino, Italia}\\
{\small $^{5}$\textit{Istituto Nazionale di Fisica Nucleare (INFN)}}\\
{\small Sezione di Torino, Via Pietro Giuria 1, 10125, Torino, Italia}}
\maketitle

\begin{abstract}
We explore the supersymmetry invariance of a supergravity theory in the
presence of a non-trivial boundary. The explicit construction of a bulk
Lagrangian based on an enlarged superalgebra, known as $AdS$-Lorentz, is
presented. Using a geometric approach we show that the supersymmetric
extension of a Gauss-Bonnet like gravity is required in order to restore the
supersymmetry invariance of the theory.
\end{abstract}

\vspace{-13.5cm}

\begin{flushright}
{\footnotesize UAI-PHY-16/11}
\end{flushright}

\vspace{11.5cm}

\section{Introduction}

The \ presence of a boundary in the context of (super)gravity has been
studied with great interest these last 40 years. In particular, the
inclusion of boundary terms plays an important role for the study of the
fruitful duality between string theory on asymptotically $AdS$ space-time
and a quantum field theory living on the boundary ($AdS$/CFT correspondence)
\cite{M1, M2, M3, M4}. The study of bulk and boundary theories has led to
the development of the so called holographic renormalization. Indeed, UV
divergences in the field theory (boundary) are related to IR divergences on
the gravitational side (bulk) which can be dealt through the holographic
renormalization procedure \cite{BK, KS, BFS}, adding appropriate
counterterms to the boundary.

At the bosonic level, the introduction of the topological Gauss-Bonnet term
to the four-dimensional $AdS$ gravity allows to regularize the action and
the related conserved charges \cite{ACOTZ, ACOTZ2, MOTZ, Olea, JKMO, JKMO2}.
Remarkably, the inclusion of the Gauss-Bonnet term does not require to
impose Dirichlet boundary conditions on the fields. On the other hand, the
addition of boundary terms to supergravity has been considered in different
approaches \cite{NV, DVB, DVBN1, DVBN2}. In particular, contrary to the
Gibbons-Hawking prescription \cite{GH}, it was pointed out that the
supergravity Lagrangian should be supersymmetric invariant without imposing
Dirichlet boundary conditions. Interestingly, it was recently shown in Ref.~%
\cite{DAA} that the introduction of a supersymmetric extension of the
Gauss-Bonnet term in a $\mathcal{N}=1$ and $\mathcal{N}=2$ supergravity
Lagrangian (with cosmological constant) allows to recover supersymmetry
invariance. \ This last result, together with the bosonic ones, suggests
that the (super)symmetry invariance of the theory requires the addition of
topological terms which besides provide the counterterms that regularize the
action.

The study of the boundary contributions needed\ to recover supersymmetry
invariance in the presence of matter or bigger supersymmetries remains
poorly explored. In this work, using a geometrical approach (rheonomic), we
explore the boundary terms needed in order to restore a particular enlarged
supersymmetry known as $AdS$-Lorentz.

The $AdS$-Lorentz (super)algebra is obtained as a deformation of the Maxwell
(super)symmetries \cite{Sorokas, DKGS}, and can be alternatively derived as
an abelian semigroup expansion ($S$-expansion) \cite{Sexp, CKMN, AMNT, ACCSP}
of the $AdS$ (super)algebra \cite{DFIMRSV, FISV, CRS, CDMR}.\ As shown in
Ref.~\cite{SS, CMR},\textbf{\ }it is possible to introduce a generalized
cosmological constant term in a Born-Infeld like gravity action when the $AdS
$-Lorentz algebra is considered. Analogously, the supersymmetric extension
of the $AdS$-Lorentz algebra allows to introduce a generalized
supersymmetric cosmological constant term in a four-dimensional supergravity
theory \cite{CRS}.

We shall first present the explicit construction of the bulk Lagrangian in
the rheonomic framework. In this geometric approach to supergravity, the
duality between a superalgebra and the Maurer-Cartan equations is used to
write down the curvatures in the superspace, whose basis is given by the
vielbein and the gravitino (bosonic and fermionic directions, respectively).
\ Subsequently, we will study the supersymmetry invariance of the Lagrangian
in the presence of a non-trivial boundary. In particular, we will show that
the supersymmetric extension of a Gauss-Bonnet like term is required in
order to restore the supersymmetry invariance of the full Lagrangian.
Interestingly, the supergravity action obtained reproduces a
MacDowell-Mansouri type action \cite{MM}.

\section{$AdS-$Lorentz Supergravity and rheonomy approach}

In the geometric framework the variational field equations obtained from the
Lagrangian are written in terms of exterior differential forms, excluding
the Hodge duality operator. Therefore they can be implemented either on the $%
x$-space manifold, or on any larger manifold containing the $x$-space. In
particular, if they are implemented on the full superspace, one obtains
algebraic relations between curvature components in $x$-space and curvature
components in directions orthogonal to $x$-space. When it happens, the
former completely determines the latter, and a solution of the field
equations on the $x$-space submanifold can be uniquely extended to a
solution of the whole group manifold. The possibility of this lifting is
called \textit{rheonomy}.

This rheonomic lifting can also be\textbf{\ }viewed as an $x$-space
transformation of the fields, which maps solutions of the $x$-space field
equations into new solutions. From this point of view, it is nothing other
than the on-shell supersymmetry transformation.

The principal demand of any supergravity theory is the invariance of the
Lagrangian under supersymmetry transformations. In the rheonomic\
(geometric) approach, the bosonic one-form $V^{a}$ $\left( a=0,1,2,3\right) $
and the fermionic one-form $\psi ^{\alpha }$ $\left( \alpha =1,\dots
,4\right) $ define the supervielbein basis in superspace \cite{CDF}. In this
framework, the supersymmetry invariance is satisfied requiring that the Lie
derivative of the Lagrangian vanishes for diffeomorphisms in the fermionic
directions of superspace,%
\begin{equation}
\delta _{\epsilon }\mathcal{L}=l_{\epsilon }\mathcal{L}=\imath _{\epsilon }d%
\mathcal{L}+d\left( \imath _{\epsilon }\mathcal{L}\right) =0\,.  \label{susy}
\end{equation}%
When a supergravity Lagrangian is considered on space-times without
boundary, the condition (\ref{susy}) trivially reduces to the first
contribution such that $\left. \imath _{\epsilon }\mathcal{L}\right\vert
_{\partial \mathcal{M}}=0$. However, in the presence of a non-trivial
boundary the condition (\ref{susy}) requires a more subtle treatment.

Before analyzing $\mathcal{N}=1$, $D=4$ $AdS$-Lorentz supergravity in the
presence of a non-trivial boundary, we will first study the construction of
the bulk Lagrangian and the corresponding supersymmetry transformation laws.
First of all, we will apply the rheonomic approach to derive the
parametrization of the $AdS$-Lorentz curvatures by studying the different
sectors of the Bianchi Identities.

\subsection{Curvatures parametrization}

The four-dimensional $AdS$-Lorentz superalgebra is generated by $\left\{
J_{ab},P_{a},Z_{ab},Q_{\alpha }\right\} $, whose generators satisfy the
(anti)commutation relations%
\begin{align}
\left[ J_{ab},J_{cd}\right] & =\eta _{bc}J_{ad}-\eta _{ac}J_{bd}-\eta
_{bd}J_{ac}+\eta _{ad}J_{bc}\,,  \label{ADSL401} \\
\left[ J_{ab},Z_{cd}\right] & =\eta _{bc}Z_{ad}-\eta _{ac}Z_{bd}-\eta
_{bd}Z_{ac}+\eta _{ad}Z_{bc}\,, \\
\left[ Z_{ab},Z_{cd}\right] & =\eta _{bc}Z_{ad}-\eta _{ac}Z_{bd}-\eta
_{bd}Z_{ac}+\eta _{ad}Z_{bc}\,,  \label{ADSL403} \\
\left[ J_{ab},P_{c}\right] & =\eta _{bc}P_{a}-\eta _{ac}P_{b}\,,\text{ \ \ \
\ }\left[ P_{a},P_{b}\right] =Z_{ab}\,, \\
\left[ Z_{ab},P_{c}\right] & =\eta _{bc}P_{a}-\eta _{ac}P_{b}\,, \\
\left[ J_{ab},Q_{\alpha }\right] & =-\frac{1}{2}\left( \gamma _{ab}Q\right)
_{\alpha }\,,\text{ \ \ \ \ }\left[ P_{a},Q_{\alpha }\right] =-\frac{1}{2}%
\left( \gamma _{a}Q\right) _{\alpha }\,, \\
\left[ Z_{ab},Q_{\alpha }\right] & =-\frac{1}{2}\left( \gamma _{ab}Q\right)
_{\alpha }\,, \\
\left\{ Q_{\alpha },Q_{\beta }\right\} & =-\frac{1}{2}\left[ \left( \gamma
^{ab}C\right) _{\alpha \beta }Z_{ab}-2\left( \gamma ^{a}C\right) _{\alpha
\beta }P_{a}\right] \,.  \label{ADSL408}
\end{align}%
Here $C$ stands for the charge conjugation matrix and $\gamma _{a}$, $\gamma
_{ab}$ are Dirac matrices. \ Let us notice that the Lorentz type algebra $%
\mathcal{L}=\left\{ J_{ab},Z_{ab}\right\} $ is a subalgebra of the above
superalgebra. This subalgebra and its extensions to higher dimensions have
been useful to derive General Relativity from Born-Infeld gravity theories
\cite{CPRS1, CPRS2, CPRS3}. Further generalizations of the $AdS$-Lorentz
superalgebra containing more than one spinor charge $Q$ can be found in Ref.~%
\cite{CRS} which can be seen as a deformation of the minimal Maxwell
superalgebras \cite{BGKL, AILW, CR1, CFRS}. Interestingly, the following
redefinition of the generators $J_{ab}\rightarrow J_{ab},$ $%
Z_{ab}\rightarrow \frac{1}{\bar{e}^{2}}Z_{ab},$ $P_{a}\rightarrow \frac{1}{%
\bar{e}}P_{a},$ $Q_{\alpha }\rightarrow \frac{1}{\bar{e}}Q_{\alpha }$
provides us with the non-standard Maxwell superalgebra in the limit $\bar{e}%
\rightarrow 0$. Let us note that the $AdS$-Lorentz superalgebra, corresponds
to a supersymmetric extension of the $\mathfrak{C}_{4}$ algebra. The $%
\mathfrak{C}_{m}$ algebras have been of particular interest in order to
derive different Lovelock gravity actions from Chern-Simons and Born-Infeld
gravity theories \cite{CMR, CDIMR}.

Let us consider the Lorentz type curvatures in the superspace which are
given by%
\begin{align}
\mathcal{R}^{ab}& =d\omega ^{ab}+\omega _{\;c}^{a}\omega ^{cb}\,,
\label{curlor1} \\
R^{a}& =D_{\omega }V^{a}+k_{\;b}^{a}V^{b}-\frac{1}{2}\overline{\psi }\gamma
^{a}\psi \,, \\
\mathcal{F}^{ab}& =D_{\omega }k^{ab}+k_{\;c}^{a}k^{cb}\,,  \label{curlor3} \\
\rho & =D_{\omega }\psi +\frac{1}{4}k^{ab}\gamma _{ab}\psi \,,
\label{curlor4}
\end{align}%
where $D_{\omega }=d+\omega $ is the Lorentz covariant exterior derivative.
They satisfy the Bianchi identities:
\begin{align}
D_{\omega }\mathcal{R}^{ab}& =0\,,  \label{eqn:drab} \\
D_{\omega }R^{a}& =\mathcal{R}_{\;b}^{a}V^{b}+\mathcal{F}_{%
\;b}^{a}V^{b}+R^{c}k_{c}^{\;a}+\overline{\psi }\gamma ^{a}\rho \,,
\label{eqn:dra} \\
D_{\omega }\mathcal{F}^{ab}& =\mathcal{R}_{\;c}^{a}k^{cb}-\mathcal{R}%
_{\;c}^{b}k^{ca}+\mathcal{F}_{\;c}^{a}k^{cb}-\mathcal{F}_{\;c}^{b}k^{ca}\,,
\label{eqn:dfab} \\
D_{\omega }\rho & =\frac{1}{4}\mathcal{R}_{ab}\gamma ^{ab}\psi +\frac{1}{4}%
\mathcal{F}_{ab}\gamma ^{ab}\psi -\frac{1}{4}k_{ab}\gamma ^{ab}\rho \,.
\label{eqn:drho}
\end{align}%
The most general Ansatz for the Lorentz type curvatures in the
super-vielbein basis $\left( V^{a},\psi \right) $ of the superspace is given
by
\begin{align}
\mathcal{R}^{ab}& =\mathcal{R}_{\;\;cd}^{ab}V^{c}V^{d}+\overline{\Theta }%
_{\;\;c}^{ab}\psi V^{c}+\alpha \bar{e}\overline{\psi }\gamma ^{ab}\psi \,,
\label{eqn:rab} \\
R^{a}& =R_{\;cd}^{a}V^{c}V^{d}+\overline{\Theta }_{\;c}^{a}\psi V^{c}+\xi
\bar{e}\overline{\psi }\gamma ^{a}\psi \,,  \label{eqn:ra} \\
\mathcal{F}^{ab}& =\mathcal{F}_{\;\;cd}^{ab}V^{c}V^{d}+\overline{\Lambda }%
_{\;\;c}^{ab}\psi V^{c}+\beta \bar{e}\overline{\psi }\gamma ^{ab}\psi \,,
\label{eqn:fab} \\
\rho & =\rho _{ab}V^{a}V^{b}+\delta \bar{e}\gamma _{a}\psi V^{a}+\Omega
_{\alpha \beta }\psi ^{\alpha }\psi ^{\beta }\,.  \label{eqn:rho}
\end{align}%
where $\bar{e}$ is the rescaling parameter. Setting $R^{a}=0$, we can
withdraw some terms appearing in the curvatures, through the study of the
scaling constraints. On the other hand, the coefficients $\alpha ,$ $\beta ,$
$\xi $ and $\delta $ appearing in the Ansatz can be determined considering
the parametrization involved in the Bianchi identities in the superspace (%
\ref{eqn:drab})-(\ref{eqn:drho}) and studying their various sectors. We
obtain that the Bianchi identities are satisfied when:%
\begin{align}
\mathcal{R}^{ab}& =\mathcal{R}_{\;\;cd}^{ab}V^{c}V^{d}+\overline{\Theta }%
_{\;\;c}^{ab}\psi V^{c}\,,  \label{eqn:parRab} \\
R^{a}& =0\,,  \label{eqn:parra} \\
\mathcal{F}^{ab}& =\mathcal{F}_{\;\;cd}^{ab}V^{c}V^{d}+\overline{\Lambda }%
_{\;\;c}^{ab}\psi V^{c}+\bar{e}\overline{\psi }\gamma ^{ab}\psi \,,
\label{eqn:parFab} \\
\rho & =\rho _{ab}V^{a}V^{b}-\bar{e}\gamma _{a}\psi V^{a}\,,
\label{eqn:parrho}
\end{align}%
where $\overline{\Theta }_{\;\;c}^{ab}=\overline{\Lambda }%
_{\;\;c}^{ab}=\epsilon ^{abde}\left( \bar{\rho}_{cd}\gamma _{e}\gamma
_{5}+\rho _{ec}\gamma _{d}\gamma _{5}-\rho _{de}\gamma _{c}\gamma
_{5}\right) $. In this way we have found the parametrization of the
curvatures and we can now consider the rheonomic construction of the bulk
Lagrangian in the geometric approach.

\subsection{Rheonomic construction of the Lagrangian}

Following the building rules for the construction of rheonomic Lagrangians
\cite{CDF}, we start by writing the most general Ansatz for the Lagrangian
as follows
\begin{equation}
\mathcal{L}=\nu ^{(4)}+F^{A}\nu _{A}^{(2)}+F^{A}F^{B}\nu _{AB}^{(0)}\,,
\label{eqn:L}
\end{equation}%
where the super-index $(p)$ denotes a $p$-form and $F^{A}$ are the super $AdS
$-Lorentz Lie algebra valued curvatures defined by%
\begin{align}
\mathcal{R}^{ab}& =d\omega ^{ab}+\omega _{\;c}^{a}\omega ^{cb}\,,
\label{curva1} \\
R^{a}& =D_{\omega }V^{a}+k_{\;b}^{a}V^{b}-\frac{1}{2}\overline{\psi }\gamma
^{a}\psi \,, \\
F^{ab}& =D_{\omega }k^{ab}+k_{\;c}^{a}k^{cb}+4\bar{e}^{2}V^{a}V^{b}+\bar{e}%
\psi \gamma ^{ab}\psi \,, \\
\Psi & =D_{\omega }\psi +\frac{1}{4}k^{ab}\gamma _{ab}\psi -\bar{e}\gamma
_{a}\psi V^{a}\,,  \label{curva4}
\end{align}%
and where
\begin{align}
& \nu ^{(4)}=\alpha _{1}\epsilon _{abcd}V^{a}V^{b}V^{c}V^{d}+\alpha _{2}%
\overline{\psi }\gamma ^{ab}\psi V^{c}V^{d}\epsilon _{abcd}+\alpha _{3}%
\overline{\psi }\gamma _{ab}\psi V^{a}V^{b}\,,  \label{eqn:Lambda} \\
&  \notag \\
& F^{A}\nu _{A}^{(2)}=\gamma _{1}\epsilon _{abcd}\mathcal{R}%
^{ab}V^{c}V^{d}+\gamma _{2}\epsilon _{abcd}F^{ab}V^{c}V^{d}+\gamma _{3}%
\overline{\Psi }\gamma _{5}\gamma _{a}\psi V^{a}+\gamma _{4}\overline{\Psi }%
\gamma _{a}\psi V^{a}+  \notag \\
& \gamma _{5}R^{a}\overline{\psi }\gamma _{a}\psi +\gamma _{6}\mathcal{R}%
^{ab}\overline{\psi }\gamma _{ab}\psi +\gamma _{7}\mathcal{R}%
^{ab}V_{a}V_{b}+\gamma _{8}\epsilon _{abcd}\mathcal{R}^{ab}\overline{\psi }%
\gamma ^{cd}\psi +  \notag \\
& +\gamma _{9}F^{ab}V_{a}V_{b}+\gamma _{10}\epsilon _{abcd}F^{ab}\overline{%
\psi }\gamma ^{cd}\psi +\gamma _{11}F^{ab}\overline{\psi }\gamma _{ab}\psi
\,,  \label{eqn:RAnuA} \\
&  \notag \\
& F^{A}F^{B}\nu _{AB}^{(0)}=\beta _{1}\mathcal{R}^{ab}\mathcal{R}_{ab}+\beta
_{2}F^{ab}F_{ab}+\beta _{3}\epsilon _{abcd}\mathcal{R}^{ab}\mathcal{R}%
^{cd}+\beta _{4}\epsilon _{abcd}\mathcal{R}^{ab}F^{cd}+  \notag \\
& +\beta _{5}\epsilon _{abcd}F^{ab}F^{cd}+\beta _{6}\overline{\Psi }\Psi
+\beta _{7}\overline{\Psi }\gamma _{5}\Psi +\beta _{8}R^{a}R_{a}\,,
\label{eqn:RARBnuAB}
\end{align}%
with $\alpha _{i},\beta _{j},\gamma _{k}$ being constants. Note that the
curvatures (\ref{curva1})-(\ref{curva4}) are invariant under the rescaling $%
\omega ^{ab}\rightarrow \omega ^{ab},$ $k^{ab}\rightarrow k^{ab},$ $%
V^{a}\rightarrow wV^{a},$ $\psi \rightarrow w^{1/2}\psi $ and $\bar{e}%
\rightarrow w^{-1}\bar{e}$. Additionally, the Lagrangian must scale with $%
w^{2}$, being $w^{2}$ the scale-weight of the Einstein term. We can prove
that the term $R^{a}R_{a}$ in (\ref{eqn:RARBnuAB}) is linear in the
curvature. Furthermore,\textbf{\ }due to scaling constraints reasons, some
of the terms in (\ref{eqn:RARBnuAB}) disappear. Here we have to observe that
a theory in $AdS$ includes a cosmological constant and, since the
coefficients appearing in the Lagrangian can be dimensional objects and
scale with negative powers of $\bar{e}$, some of the terms in $F^{A}F^{B}\nu
_{AB}^{(0)}$ can survive the scaling and contribute to the Lagrangian as
total derivatives. However, since we are now constructing the bulk
Lagrangian, we can neglect them and set $F^{A}F^{B}\nu _{AB}^{(0)}=0$. We
will show that these terms will be fundamental for the construction of the
boundary Lagrangian.

Let us consider now the scaling in (\ref{eqn:Lambda}) whose coefficients
must be redefined in the following way in order to give non-vanishing
contributions to the Lagrangian:
\begin{equation}
\alpha _{1}\equiv \bar{e}^{2}\alpha _{1}^{\prime }\,,\text{ \ \ }\alpha
_{2}\equiv \bar{e}\alpha _{2}^{\prime },\text{ \ \ }\alpha _{3}\equiv \bar{e}%
\alpha _{3}^{\prime }\,.
\end{equation}%
In this way, all the terms in $\nu $ scale as $w^{2}$. Then, applying the
scaling and the parity conservation law to (\ref{eqn:Lambda}) and (\ref%
{eqn:RAnuA}) we obtain
\begin{equation}
\alpha _{3}=0\,;\ \text{\ \ \ \ }\gamma _{4}=\gamma _{5}=\gamma _{6}=\gamma
_{7}=\gamma _{8}=\gamma _{9}=\gamma _{10}=\gamma _{11}=0\,.
\end{equation}

Therefore, we are left with the Lagrangian
\begin{equation}
\mathcal{L}=\epsilon _{abcd}\mathcal{R}^{ab}V^{c}V^{d}+\gamma _{3}\overline{%
\psi }\gamma _{a}\gamma _{5}\Psi V^{a}+\gamma _{2}\epsilon
_{abcd}F^{ab}V^{c}V^{d} \\
+\alpha _{1}^{\prime }\bar{e}^{2}\epsilon _{abcd}V^{a}V^{b}V^{c}V^{d}+\alpha
_{2}^{\prime }\bar{e}\epsilon _{abcd}\overline{\psi }\gamma ^{ab}\psi
V^{c}V^{d}\,,
\end{equation}%
where we have consistently set $\gamma _{1}=1$. Using the definition of the $%
AdS$-Lorentz curvatures (\ref{curva1})-(\ref{curva4}), we can write%
\begin{align*}
\mathcal{L}& =\epsilon _{abcd}\mathcal{R}^{ab}V^{c}V^{d}+\gamma _{3}%
\overline{\psi }\gamma _{a}\gamma _{5}D_{\omega }\psi V^{a}+\frac{\gamma _{3}%
}{4}\epsilon _{abcd}k^{ab}\overline{\psi }\gamma ^{c}\psi V^{d} \\
& +\gamma _{2}\epsilon _{abcd}\left( D_{\omega
}k^{ab}+k_{\;c}^{a}k^{cb}\right) V^{c}V^{d}+\left( \alpha _{1}^{\prime
}+4\gamma _{2}\right) \bar{e}^{2}\epsilon _{abcd}V^{a}V^{b}V^{c}V^{d} \\
& +\left( \alpha _{2}^{\prime }+\gamma _{2}+\frac{\gamma _{3}}{2}\right)
\bar{e}\epsilon _{abcd}\overline{\psi }\gamma ^{ab}\psi V^{c}V^{d}\,.
\end{align*}%
We can now determine the coefficients $\alpha _{1}^{\prime },$ $\alpha
_{2}^{\prime },$ $\gamma _{2}$ and $\gamma _{3}$ through the study of the
field equations. In order to obtain them, let us compute the variation of
the Lagrangian with respect to the different fields. The variation of the
Lagrangian with respect to the spin connection $\omega ^{ab} $ is given by%
\begin{equation}
\delta _{\omega }\mathcal{L}=2\epsilon _{abcd}\delta \omega ^{ab}\left(
D_{\omega }V^{c}+\gamma _{2}k_{\;f}^{c}V^{f}-\frac{1}{8}\gamma _{3}\overline{%
\psi }\gamma ^{c}\psi \right) V^{d}\,.
\end{equation}%
Here we see that, if $\gamma _{2}=1$ and $\gamma _{3}=4$, $\delta _{\omega }%
\mathcal{L}=0$ leads to the field equation for the $AdS$-Lorentz
supertorsion:
\begin{equation}
\epsilon _{abcd}R^{c}V^{d}=0\,.
\end{equation}%
The variation of the Lagrangian with respect to $k^{ab}$ gives the same
result. \

On the other hand, the variation of the Lagrangian with respect to the
vielbein $V^{a}$ leads to
\begin{equation}
2\epsilon _{abcd}(\mathcal{R}^{ab}V^{c}+F^{ab}V^{c})+4\overline{\psi }\gamma
_{d}\gamma _{5}\Psi =0\,,
\end{equation}%
where we have used
\begin{equation*}
\epsilon _{abcd}k^{ab}\overline{\psi }\gamma ^{c}\psi =\overline{\psi }%
\gamma _{d}\gamma _{5}k^{ab}\gamma _{ab}\psi \,,
\end{equation*}%
and where we have set $\alpha _{1}^{\prime }=-2$ and $\alpha _{2}^{\prime
}=-1$, in order to recover the $AdS$-Lorentz curvatures. In the same way,
from the variation with respect to the gravitino field $\psi $ we find the
following field equation:%
\begin{equation}
8V^{a}\gamma _{a}\gamma _{5}\Psi +4\gamma _{a}\gamma _{5}\psi R^{a}=0\,.
\end{equation}%
Summarizing, we have found the following values for the coefficients:
\begin{equation}
\alpha _{1}^{\prime }=-2,\text{ \ \ }\alpha _{2}^{\prime }=-1,\text{ \ \ }%
\gamma _{2}=1,\text{ \ \ }\gamma _{3}=4\,.
\end{equation}%
Thus we have completely determined the bulk Lagrangian $\mathcal{L}_{bulk}$
of the theory, which can be written in terms of the Lorentz type curvatures (%
\ref{curlor1})-(\ref{curlor4}) as follows
\begin{align}
\mathcal{L}_{bulk}& =\epsilon _{abcd}\mathcal{R}^{ab}V^{c}V^{d}+\epsilon
_{abcd}\mathcal{F}^{ab}V^{c}V^{d}+4\overline{\psi }\gamma _{a}\gamma
_{5}\rho V^{a}  \notag \\
& +2\bar{e}^{2}\epsilon _{abcd}V^{a}V^{b}V^{c}V^{d}+2\bar{e}\epsilon _{abcd}%
\overline{\psi }\gamma ^{ab}\psi V^{c}V^{d}\,.
\end{align}

\subsection{Supersymmetry transformation laws}

The parametrizations we got in the previous section allow to obtain the
supersymmetry transformation laws. \ Indeed, in the rheonomic formalism, the
transformations on space-time are given by%
\begin{equation}
\delta \mu ^{A}=\left( \nabla \epsilon \right) ^{A}+l_{\epsilon }F^{A}\,,
\end{equation}%
where $\epsilon ^{A}\equiv \left( \epsilon ^{ab},\epsilon ^{a},\varepsilon
^{ab},\epsilon \right) $. Then, restricting us to supersymmetric
transformations we have $\epsilon ^{ab}=\epsilon ^{a}=\varepsilon ^{ab}=0$
and%
\begin{align}
& l_{\epsilon }(\mathcal{R}^{ab})=\overline{\Theta }_{\;\;c}^{ab}\epsilon
V^{c}\,,  \label{eqn:susy} \\
& l_{\epsilon }(R^{a})=0\,, \\
& l_{\epsilon }(\mathcal{F}^{ab})=\overline{\Lambda }_{\;\;c}^{ab}\epsilon
V^{c}+2\bar{e}\overline{\epsilon }\gamma ^{ab}\psi \,, \\
& l_{\epsilon }(\rho )=-\bar{e}\gamma _{a}\epsilon V^{a}\,,
\end{align}%
which provide the following supersymmetry transformation laws:%
\begin{eqnarray*}
\delta _{\epsilon }\omega ^{ab} &=&\overline{\Theta }_{\;\;c}^{ab}\epsilon
V^{c}\,, \\
\delta _{\epsilon }V^{a} &=&\bar{\epsilon}\gamma ^{a}\psi \text{\thinspace },
\\
\delta _{\epsilon }k^{ab} &=&-2\bar{e}\overline{\epsilon }\gamma ^{ab}\psi +%
\overline{\Lambda }_{\;\;c}^{ab}\epsilon V^{c}\,, \\
\delta _{\epsilon }\psi  &=&d\epsilon +\frac{1}{4}\omega ^{ab}\gamma
_{ab}\epsilon +\frac{1}{4}k^{ab}\gamma _{ab}\epsilon +\bar{e}\gamma
_{a}\epsilon V^{a}\,.
\end{eqnarray*}%
Under these transformation laws the Lagrangian is invariant up to boundary
terms. \ The presence of a boundary requires to check explicitly the
condition (\ref{susy}).

\section{Supersymmetry invariance in the presence of a boundary}

In this section, following the approach presented in Ref.~\cite{DAA}, we
analyze the supersymmetry invariance of the Lagrangian in the presence of a
non-trivial boundary. \ In particular, we present the explicit boundary
terms required in order to recover the full supersymmetry invariance of the
Lagrangian.

Let us consider the Lagrangian found in the previous section,%
\begin{align}
\mathcal{L}_{bulk}& =\epsilon _{abcd}\mathcal{R}^{ab}V^{c}V^{d}+4\bar{\psi}%
V^{a}\gamma _{a}\gamma _{5}\rho   \notag \\
& +\epsilon _{abcd}\left( \mathcal{F}^{ab}V^{c}V^{d}+2\bar{e}V^{a}V^{b}\bar{%
\psi}\gamma ^{cd}\psi +2\bar{e}^{2}V^{a}V^{b}V^{c}V^{d}\right) \,.
\end{align}%
The supersymmetry invariance in the bulk is satisfied on-shell%
\begin{equation*}
R^{a}=0\,.
\end{equation*}%
Nevertheless, the boundary invariance of the Lagrangian under supersymmetry
is not trivially satisfied:%
\begin{equation}
l_{\epsilon }\mathcal{L}_{bulk}\mathcal{|}_{\partial \mathcal{M}_{4}}\neq
0\,.
\end{equation}%
In order to recover the supersymmetric invariance of the theory, we require
a more subtle approach. Indeed, we have to add boundary terms to the bulk
Lagrangian.

The only boundary contributions compatible with parity, Lorentz-like
invariance and $\mathcal{N}=1$ supersymmetry are%
\begin{eqnarray*}
d\left( \varpi ^{ab}\mathcal{N}^{cd}+\varpi _{\text{ }f}^{a}\varpi
^{fb}\varpi ^{cd}\right) \epsilon _{abcd} &=&\epsilon _{abcd}\mathcal{N}^{ab}%
\mathcal{N}^{cd}\,, \\
d\left( \bar{\rho}\gamma _{5}\psi \right)  &=&\bar{\rho}\gamma _{5}\rho +%
\frac{1}{8}\epsilon _{abcd}\mathcal{N}^{ab}\bar{\psi}\gamma ^{cd}\psi \,,
\end{eqnarray*}%
where we have defined $\varpi ^{ab}=\omega ^{ab}+k^{ab}$ and $\mathcal{N}%
^{ab}=\mathcal{R}^{ab}+\mathcal{F}^{ab}$, with $\mathcal{R}^{ab}$ and $%
\mathcal{F}^{ab}$ given by eqs. (\ref{curlor1}) and (\ref{curlor3}),
respectively. One can notice that $\varpi ^{ab}$ and $\mathcal{N}^{ab}$ are
related to a Lorentz-like generator $M_{ab}=J_{ab}+Z_{ab}$ (see eqs. (\ref%
{ADSL401}) - (\ref{ADSL403})). Thus, let us consider the following boundary
Lagrangian%
\begin{align}
\mathcal{L}_{bdy}& =\alpha \epsilon _{abcd}\left( \mathcal{R}^{ab}\mathcal{R}%
^{cd}+2\epsilon _{abcd}\mathcal{R}^{ab}\mathcal{F}^{cd}+\epsilon _{abcd}%
\mathcal{F}^{ab}\mathcal{F}^{cd}\right)   \notag \\
& +\beta \left( \bar{\rho}\gamma _{5}\rho +\frac{1}{8}\epsilon _{abcd}%
\mathcal{R}^{ab}\bar{\psi}\gamma ^{cd}\psi +\frac{1}{8}\epsilon _{abcd}%
\mathcal{F}^{ab}\bar{\psi}\gamma ^{cd}\psi \right) \,.
\end{align}%
Let us note that the structure of a supersymmetric Gauss-Bonnet like gravity
appears.\ Then, the full Lagrangian is given by%
\begin{align}
\mathcal{L}_{full}& =\mathcal{L}_{bulk}+\mathcal{L}_{bdy}  \notag \\
& =\epsilon _{abcd}\mathcal{R}^{ab}V^{c}V^{d}+4\bar{\psi}V^{a}\gamma
_{a}\gamma _{5}\rho +\epsilon _{abcd}\left( \mathcal{F}^{ab}V^{c}V^{d}+2\bar{%
e}V^{a}V^{b}\bar{\psi}\gamma ^{cd}\psi +2\bar{e}^{2}V^{a}V^{b}V^{c}V^{d}%
\right)   \notag \\
& +\alpha \epsilon _{abcd}\left( \mathcal{R}^{ab}\mathcal{R}^{cd}+2\epsilon
_{abcd}\mathcal{R}^{ab}\mathcal{F}^{cd}+\epsilon _{abcd}\mathcal{F}^{ab}%
\mathcal{F}^{cd}\right)   \notag \\
& +\beta \left( \frac{1}{8}\epsilon _{abcd}\mathcal{R}^{ab}\bar{\psi}\gamma
^{cd}\psi +\frac{1}{8}\epsilon _{abcd}\mathcal{F}^{ab}\bar{\psi}\gamma
^{cd}\psi +\bar{\rho}\gamma _{5}\rho \right) \,.
\end{align}%
Due to the $\bar{e}^{-2}$-homogeneous scaling of the Lagrangian, we have
that the coefficients $\alpha $ and $\beta $ must be related to $\bar{e}^{-2}
$ and $\bar{e}^{-1}$, respectively.

As we have previously pointed out, the supersymmetry invariance of the full
Lagrangian $\mathcal{L}_{full}$ requires the following condition%
\begin{equation}
\delta _{\epsilon }\mathcal{L}_{full}=l_{\epsilon }\mathcal{L}_{full}=\imath
_{\epsilon }d\mathcal{L}_{full}+d\left( \imath _{\epsilon }\mathcal{L}%
_{full}\right) =0\,.
\end{equation}%
Naturally, the condition for supersymmetry in the bulk $\imath _{\epsilon }d%
\mathcal{L}_{full}=0$ is satisfied since the boundary contributions
correspond to total derivatives. \ Thus the supersymmetry invariance of the
full Lagrangian $\mathcal{L}_{full}$ requires to verify the condition $%
\imath _{\epsilon }\left( \mathcal{L}_{full}\right) =0$ on the boundary. In
particular, we have%
\begin{align}
\imath _{\epsilon }\left( \mathcal{L}_{full}\right) & =\epsilon
_{abcd}\imath _{\epsilon }\left( \mathcal{R}^{ab}+\mathcal{F}^{ab}\right)
V^{c}V^{d}+4\bar{\epsilon}V^{a}\gamma _{a}\gamma _{5}\rho +4\bar{\psi}%
V^{a}\gamma _{a}\gamma _{5}\imath _{\epsilon }\left( \rho \right)   \notag \\
& +\epsilon _{abcd}4\bar{e}V^{a}V^{b}\bar{\epsilon}\gamma ^{cd}\psi +2\imath
_{\epsilon }\left( \mathcal{R}^{ab}+\mathcal{F}^{ab}\right) \left\{ \alpha
\mathcal{R}^{cd}+\frac{\beta }{16}\bar{\psi}\gamma ^{cd}\psi +\alpha
\mathcal{F}^{cd}\right\} \epsilon _{abcd}  \notag \\
& +\frac{\beta }{4}\epsilon _{abcd}\left( \mathcal{R}^{ab}+\mathcal{F}%
^{ab}\right) \bar{\epsilon}\gamma ^{cd}\psi +2\beta \imath _{\epsilon
}\left( \bar{\rho}\right) \gamma _{5}\rho \,.
\end{align}%
Then, $\left. \frac{\delta \mathcal{L}_{full}}{\delta \mu ^{A}}\right\vert
_{\partial \mathcal{M}}=0$ implies the following constraints on the boundary:%
\begin{eqnarray}
\left( \mathcal{R}^{ab}+\mathcal{F}^{ab}\right) |_{\partial \mathcal{M}} &=&-%
\frac{1}{2\alpha }V^{a}V^{b}-\frac{\beta }{16\alpha }\bar{\psi}\gamma
^{ab}\psi \,,\text{ } \\
\rho |_{\partial \mathcal{M}} &=&\frac{2}{\beta }V^{a}\gamma _{a}\psi \,.
\end{eqnarray}%
The supersymmetry invariance requires $\imath _{\epsilon }\left( \mathcal{L}%
_{full}\right) |_{\partial \mathcal{M}}=0$. \ Thus we find

\begin{align*}
\imath _{\epsilon }\left( \mathcal{L}_{full}\right) |_{\partial \mathcal{M}%
}& =-\frac{\beta }{8\alpha }\epsilon _{abcd}\bar{\epsilon}\gamma ^{ab}\psi
V^{c}V^{d}+4\bar{\epsilon}V^{a}\gamma _{a}\gamma _{5}\rho +\frac{8}{\beta }%
\bar{\psi}V^{a}\gamma _{a}\gamma _{5}V^{b}\gamma _{b}\epsilon  \\
& +4\bar{e}\epsilon _{abcd}V^{a}V^{b}\bar{\epsilon}\gamma ^{cd}\psi -\left(
\frac{\beta }{4\alpha }\bar{\epsilon}\gamma ^{ab}\psi \right) \left\{ \alpha
\mathcal{R}^{cd}+\frac{\beta }{16}\bar{\psi}\gamma ^{cd}\psi +\alpha
\mathcal{F}^{cd}\right\} \epsilon _{abcd} \\
& +\frac{\beta }{4}\epsilon _{abcd}\left\{ \mathcal{R}^{ab}\bar{\epsilon}%
\gamma ^{cd}\psi +\mathcal{F}^{ab}\bar{\epsilon}\gamma ^{cd}\psi \right\} -4%
\bar{\epsilon}\gamma _{a}V^{a}\gamma _{5}\rho \,.
\end{align*}%
Using the Fierz identity for $\mathcal{N}=1$, $\gamma _{ab}\psi \bar{\psi}%
\gamma ^{ab}\psi =0$, we have%
\begin{equation*}
\imath _{\epsilon }\left( \mathcal{L}_{full}\right) |_{\partial \mathcal{M}%
}=\left( 4\bar{e}-\frac{\beta }{8\alpha }\right) \epsilon _{abcd}\bar{%
\epsilon}\gamma ^{ab}\psi V^{c}V^{d}+\frac{8}{\beta }\bar{\psi}V^{a}\gamma
_{a}\gamma _{5}V^{b}\gamma _{b}\epsilon \,.
\end{equation*}%
Then, using the gamma matrices identity, we have that the supersymmetry
invariance implies the following relation between $\alpha $ and $\beta $:%
\begin{equation}
\frac{\beta }{4\alpha }+\frac{8}{\beta }=8\bar{e}\,.
\end{equation}%
Solving for $\beta $ we find%
\begin{equation}
\beta =16e\alpha \left( 1\pm \sqrt{1-\frac{1}{8\bar{e}^{2}\alpha }}\right)
\,.
\end{equation}%
Let us note that the root vanishes for%
\begin{equation*}
\alpha =\frac{1}{8\bar{e}^{2}}\,,
\end{equation*}%
which implies
\begin{equation*}
\beta =\frac{2}{\bar{e}}\,.
\end{equation*}%
Interestingly, with these values for $\alpha $ and $\beta $ we recover the
following $2$-form curvatures%
\begin{eqnarray}
N^{ab} &=&\mathcal{R}^{ab}+\mathcal{F}^{ab}+4\bar{e}^{2}V^{a}V^{b}+\bar{e}%
\bar{\psi}\gamma ^{ab}\psi \,,  \label{2f1} \\
\Psi  &=&\rho -\bar{e}V^{a}\gamma _{a}\psi \,\,,  \label{2f2} \\
R^{a} &=&D_{\omega }V^{a}+k_{\;b}^{a}V^{b}-\frac{1}{2}\bar{\psi}\gamma
^{a}\psi \,.
\end{eqnarray}%
which reproduce the $AdS$-Lorentz curvatures with%
\begin{eqnarray*}
\,N^{ab} &=&\mathcal{R}^{ab}+F^{ab}\,,\text{ \ \ \ where} \\
\mathcal{R}^{ab} &=&d\omega ^{ab}+\omega _{\text{ }c}^{a}\omega ^{cb}\,, \\
F^{ab} &=&\mathcal{F}^{ab}+4\bar{e}^{2}V^{a}V^{b}+\bar{e}\bar{\psi}\gamma
^{ab}\psi \,.
\end{eqnarray*}%
Finally, the full Lagrangian can be written as a MacDowell-Mansouri like
form in terms of the $2$-form curvatures (\ref{2f1}) - (\ref{2f2}),%
\begin{equation}
\mathcal{L}_{full}=\frac{1}{8\bar{e}^{2}}\epsilon _{abcd}N^{ab}N^{cd}+\frac{2%
}{\bar{e}}\bar{\Psi}\gamma _{5}\Psi \,,  \label{LFULL}
\end{equation}%
whose boundary term corresponds to a supersymmetric Gauss-Bonnet like term,

\begin{equation}
\mathcal{L}_{bdy}=\frac{1}{8\bar{e}^{2}}\epsilon _{abcd}\left( \mathcal{R}%
^{ab}\mathcal{R}^{cd}+2\mathcal{R}^{ab}\mathcal{F}^{cd}+\mathcal{F}^{ab}%
\mathcal{F}^{cd}\right) +\frac{4}{\bar{e}}\left( \frac{1}{8}\epsilon _{abcd}%
\mathcal{R}^{ab}\bar{\psi}\gamma ^{cd}\psi +\frac{1}{8}\epsilon _{abcd}%
\mathcal{F}^{ab}\bar{\psi}\gamma ^{cd}\psi +\bar{\rho}\gamma _{5}\rho
\right).
\end{equation}

This term allows to recover the supersymmetric invariance of the theory in
the presence of a boundary. \ The same phenomenon occurs in pure gravity,
where the Gauss-Bonnet term assures the invariance of the Lagrangian in the
presence of a non-trivial boundary. Additionally, the supersymmetric
extension of the Gauss-Bonnet term was introduced in Ref. \cite{DAA}, in
order to restore the supersymmetry invariance in $\mathcal{N}=1$ and $%
\mathcal{N}=2$, $Osp\left( \mathcal{N}|4\right) $ supergravity in the
presence of a boundary.

On the other hand, the bulk Lagrangian reproduces the generalized
supersymmetric cosmological term presented in Ref.~\cite{CRS}, and
corresponds to a supersymmetric extension of the results found in Refs.~\cite%
{SS, AKL}.

Let us note that an In\"{o}n\"{u}-Wigner (IW) contraction of the full
Lagrangian (\ref{LFULL}) leads to the Maxwell MacDowell-Mansouri Lagrangian
presented in Ref.~\cite{CR2}, corresponding to $\mathcal{N}=1$ pure
supergravity Lagrangian in the presence of a non-trivial boundary.

\section{Comments and possible developments}

In this paper we have first of all presented the explicit construction of
the $\mathcal{N}=1$, $D=4$ $AdS$-Lorentz supergravity bulk Lagragian in the
rheonomic framework. In particular, we have shown an alternative way to
introduce a generalized supersymmetric cosmological term to supergravity. \
Subsequently, we have studied the supersymmetry invariance of the Lagrangian
in the presence of a non-trivial boundary. Interestingly, the supersymmetric
extension of a Gauss-Bonnet like term is required in order to restore the
supersymmetry invariance of the full Lagrangian. The addition of a
topological boundary term in a four-dimensional bosonic action is equivalent
to the holographic renormalization in the $AdS/$CFT formalism. \ Then, it
seems that the presence of the $k^{ab}$ fields through the $\mathcal{F}^{ab}$
curvature in the boundary would allow to regularize the supergravity action
in the holographic renormalization language. Additionally, as was pointed
out in Refs.~\cite{MO, MTO}, the bosonic MacDowell-Mansouri action is
on-shell equivalent to the square of the Weyl tensor describing conformal
gravity. \ Thus, the supergravity action \`{a} la MacDowell-Mansouri would
suggest a superconformal structure which represents an additional motivation
in our approach.

The results obtained here could be useful in order to study supergravity
theories in the presence of a non-trivial boundary in higher dimensions or
coupled to matter. In particular, it would be interesting to analyze the
boundary terms necessary to restore the supersymmetry invariance of a
general matter coupled $\mathcal{N}=2$ supergravity considering the bulk
Lagrangians introduced in Refs.~\cite{ABCAFFM, ACDRT}.

\section{Acknowledgment}

This work was supported in part by FONDECYT Grants No 1130653 (MCI) and also
by the Newton-Picarte CONICYT Grant No. DPI20140053 (P.K.C. and E.K.R.). \
MCI was supported by grants from CONICYT and from the Universidad de Concepci%
\'{o}n, Chile. \ The authors wish to thank L. Andrianopoli, R. D'Auria and
M. Trigiante for enlightening discussions.

\end{document}